\documentclass{optica-article}

\usepackage{opticajournal} % for journals or Optica Open

\articletype{Research Article}
\usepackage{lineno}
\usepackage{makecell}
\usepackage{svg}
\usepackage{physics}
\usepackage{multirow}
\usepackage{nicematrix}
\usepackage{xcolor}
\usepackage{multirow}
\usepackage{soul}
\sethlcolor{yellow}

\begin{document}
\title{Spontaneous Brillouin Scattering in a Few-Mode Optical Fiber}
\author{Hikari Kikuchi,\authormark{1} Rekishu Yamazaki\authormark{1*}}
\address{\authormark{1} Department of Natural Sciences, International Christian University, Mitaka, Tokyo 181-8585, Japan\\
}
\email{\authormark{*}rekishu@icu.ac.jp} 

\begin{abstract*} 
We report a comprehensive experimental study of spontaneous Brillouin scattering in a few-mode optical fiber, resolving both forward and backward scattering processes for intra- and inter-modal interactions. Using heterodyne detection, Stokes and anti-Stokes components without external acoustic excitation are observed and quantitatively extracted Brillouin shifts, linewidths, and gain coefficients. Forward scattering is mediated by guided torsional–radial acoustic modes with frequencies ranging from MHz to GHz, while backward scattering involves longitudinal core-guided modes at frequencies of tens of GHz. These results provide calibrated benchmarks for Brillouin interactions in few-mode fibers, offering insights relevant to phonon-based quantum applications and mode-selective optomechanics.
\end{abstract*}

%%%%%%%%%%%%%%%%%%%%%%%%%%  body  %%%%%%%%%%%%%%%%%%%%%%%%%%
\section{Introduction}
 
Brillouin scattering is a nonlinear interaction between light and acoustic phonons that produces frequency-shifted scattered photons~\cite{boyd2020,merklein2022100}. Coupling between optical and mechanical modes enables new functionalities such as quantum transduction\cite{safavi2011proposal}, sensing\cite{foreman2015whispering}, and information storage\cite{riedinger2018remote,merklein2018brillouin}.  In optical fibers, Brillouin interactions have been extensively explored and form the basis for applications such as sensing \cite{denisov2016going}, narrow-linewidth lasers \cite{smith1991narrow,bai2022comprehensive,bashan2022forward,corvo1988forward}, and slow light \cite{okawachi2005tunable}. 
In single-mode optical fibers, Brillouin scattering is well understood and typically occurs in a backward intra-modal configuration, where the optical pump and scattered wave propagate counter-linearly in the same spatial mode (usually $LP_{01}$), coupled via a longitudinal acoustic wave at GHz-tens of GHz frequencies~\cite{yeniay2002spontaneous}. In contrast, multimode and few-mode fibers support multiple optical and acoustic modes, enabling the observation of intricate inter-modal Brillouin interactions~\cite{song2013characterization}. However, while backward Brillouin scattering has been extensively studied and applied, measuring forward Brillouin scattering remains challenging.  

In forward scattering, both the pump and scattered waves propagate co-linearly, and only the acoustic waves with a very small wavenumber can satisfy the phase-matching condition. The resulting acoustic modes typically have  MHz–hundreds of MHz frequencies and often appear to have a phase velocity exceeding the speed of light~\cite{1130012844551500709,wolff2021brillouin}. Detection of the forward scattering is challenging due to weak signals and strong spatial overlap with the pump beam~\cite{sanchez2022recent}. Moreover, the linewidth of the forward Brillouin signal is significantly narrower than that of backward scattering, often on the order of 10 to 100 kHz, reflecting the long phonon lifetimes of the guided acoustic modes involved~\cite{1130012844551500709}. 

Forward Brillouin scattering in few-mode fibers opens up unique opportunities for mode-selective excitation of specific acoustic modes~\cite{diamandi2022interpolarization}, particularly through inter-modal forward Brillouin scattering. However, experimental studies remain scarce, and detection techniques are largely limited to stimulated configurations~\cite{HarnessingIntra-Mode, layosh2025forward}. In this article, we present a comprehensive study of spontaneous Brillouin scattering in a few-mode fiber, where both forward- and backward-scattered signals are observed for intra- and inter-modal processes and are analyzed in comparison with theoretical predictions. This work presents the first experimental observation of both Stokes and anti-Stokes components in forward Brillouin scattering, and, in contrast to earlier reports that were limited to relative gain measurements \cite{layosh2025forward}, yields a calibrated absolute Brillouin gain.

\section{Theory}
\subsection{Dispersion Relations and Phase Matching Condition}

A large Brillouin signal can be observed when two guided optical modes and an acoustic mode satisfy the phase-matching condition~\cite{Shen1965, Smith1972}.  Phase matching corresponds to the conservation of energy and momentum in Brillouin scattering and can be conveniently visualized using the dispersion relations of optical and acoustic modes.

\begin{figure}[htbp]
  \centering
  \includegraphics[width=0.9\linewidth]{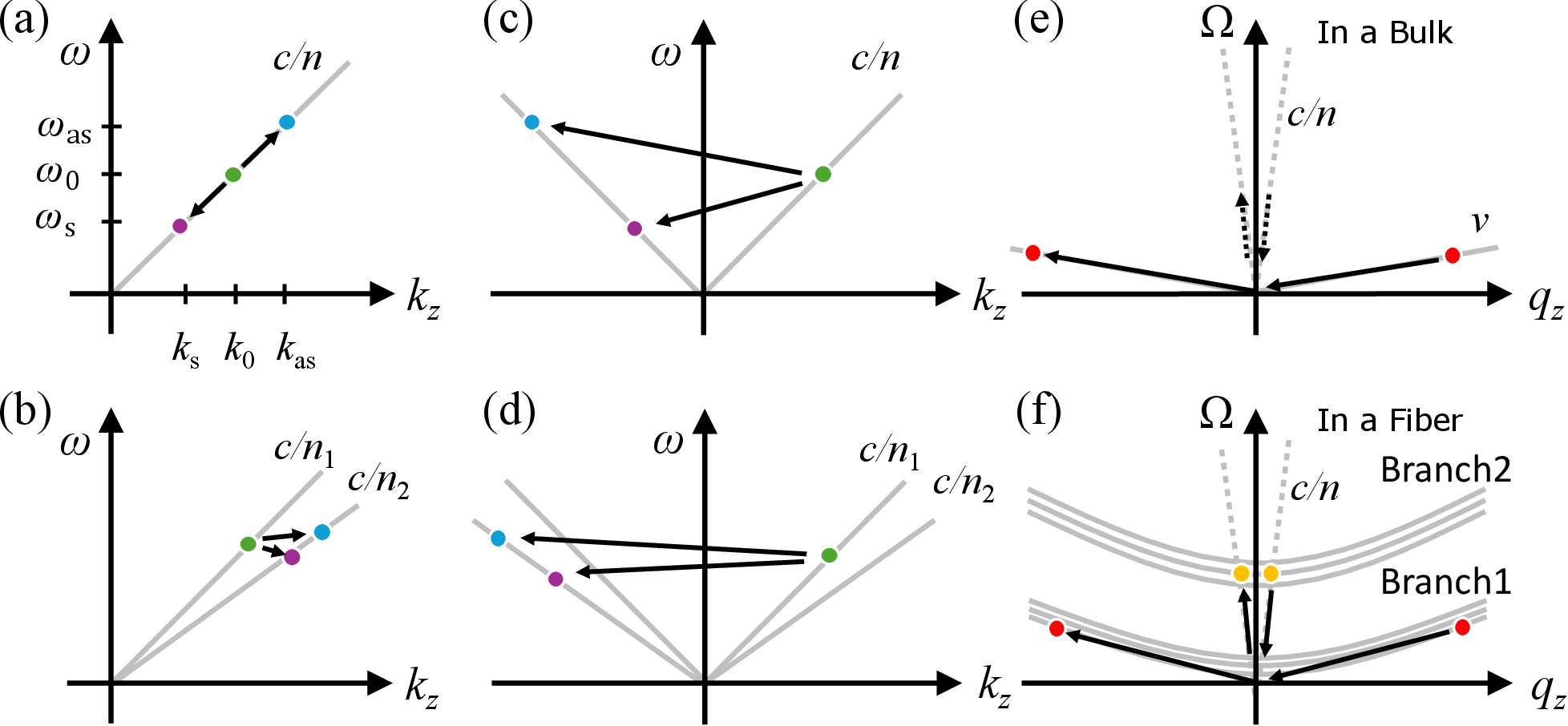} 
  \caption{Optical and acoustic dispersion relations for various Brillouin scattering configurations. (a) Forward scattering within a single optical mode (intra-modal scattering). An initial optical state (green) with frequency $\omega_0$ and wavenumber $k_0$ is scattered to generate an anti-Stokes signal (blue) and a Stokes signal (magenta). (b) Forward scattering involving two optical modes with refractive indices $n_1$ and $n_2$. (c) Backward scattering within a single optical mode, where a large momentum transfer is required, in contrast to forward scattering. (d) Backward scattering between two optical modes with refractive indices $n_1$ and $n_2$. (e) Acoustic dispersion relation for a bulk acoustic mode. (f) Acoustic dispersion relations in an optical fiber, showing multiple guided acoustic branches.}
  \label{fig:disp}
\end{figure}

Figure~\ref{fig:disp} illustrates the optical dispersion relations for (a) intra-modal forward, (b) inter-modal forward, (c) intra-modal backward, and (d) inter-modal backward Brillouin scattering. In each case, an incident optical mode with frequency $\omega_0$ and wavenumber $k_0$ is scattered into anti-Stokes and Stokes waves with frequencies $\omega_{\mathrm{as}}$, $\omega_{\mathrm{s}}$ and wavenumbers $k_{\mathrm{as}}$, $k_{\mathrm{s}}$, respectively. Different optical modes generally have distinct effective refractive indices, denoted here as $n_1$ and $n_2$, leading to different dispersion relations. Phase matching requires the acoustic frequency and wavenumber to satisfy the conditions $\Omega = |\omega_0-\omega_\text{as,s}|$ and $|q_z| = |k_0-k_\text{as,s}|$, respectively.
A key distinction between forward and backward scattering is the momentum transfer involved. Backward scattering requires a large axial acoustic wavenumber, whereas forward scattering occurs with a small wavenumber mismatch.

Figures~\ref{fig:disp}(e) and \ref{fig:disp}(f) show acoustic dispersion relations for a bulk medium and for an optical fiber, respectively. Since the acoustic velocity is much slower than that of light ($v\ll c/n$), in bulk media, one can only find an acoustic mode for the backward scattering, denoted with a red dots. In contrast, transverse confinement in optical fibers leads to the emergence of ``optical-like'' acoustic modes with finite cutoff frequencies at $q_z=0$. These acoustic modes, denoted with orange dots, allow the forward scattering.  Additionally, multiple acoustic branches corresponding to longitudinal and shear modes are observed in the present experiment.

Depending on the position along the dispersion curve, acoustic modes exhibit different characteristics. Optical-like phonons near $q_z=0$ possess small axial wavenumbers and finite frequencies, resulting in large phase velocities that may exceed the speed of light. In contrast, acoustic-like phonons at large $q_z$ have phase velocities comparable to those of bulk acoustic waves. Forward Brillouin scattering predominantly involves phonons near $q_z=0$, whereas backward scattering is mediated by phonons with large axial wavenumbers.

\subsection{Brillouin shift for various scattering}

Brillouin frequency shifts $\Omega$ for different scattering configurations are summarized in Table~\ref{tab:1}. 

\begin{table}[htbp]
    \centering
    \caption{Brillouin shift for various scattering processes}
    \label{tab:1}
    \begin{NiceTabular}{ccccc}
        \hline
        \makecell{Scattering \\ Direction}  & Optical Mode & \makecell{Anti-Stokes \\ Process}  & \makecell{Stokes \\ Process} & Other Condition \\
        \hline
        \Block{2-1}{Backward} & Intra-mode & \multicolumn{1}{l}{$\Omega_\text{as}=\frac{2v}{c/n-v}\omega_0$} & \multicolumn{1}{l}{$\Omega_\text{s}=\frac{2v}{c/n+v}\omega_0$} &  \\ \cline{2-5}
                     & Inter-mode & \multicolumn{1}{l}{$\Omega_\text{as}=\frac{(n_1+n_2)v\omega_0}{c-n_2v}$} & \multicolumn{1}{l}{$\Omega_\text{s}=\frac{(n_1+n_2)v\omega_0}{c+n_2v}$} &  \\\hline
        \Block{2-1}{Forward} & Intra-mode & \Block{2-2}{\multicolumn{1}{l}{$\Omega_\text{as} = \Omega_{s}\approx  \frac{2l+4m-1}{4}\frac{\pi v}{R}$     }} & & \Block{2-1}{$v=\frac{\Omega}{q_z}=\frac{c}{n}$} \\\cline{2-2}
                     & Inter-mode &                                  & & \\
        \hline
    \end{NiceTabular}
\end{table}

In backward Brillouin scattering, the frequency shift is typically on the order of tens of GHz, reflecting the large momentum transfer required. In contrast, forward Brillouin scattering occurs at much lower frequencies, typically ranging from a few MHz to several hundred MHz.
Forward Brillouin scattering requires the phase velocity of the acoustic mode to match the optical group velocity, expressed as $v = \Omega/q_z = c/n$.
While this condition cannot be satisfied by bulk acoustic modes, it can be fulfilled by optical-like guided acoustic modes possessing finite cutoff frequencies. As indicated by the orange dot in Fig.~\ref{fig:disp}(f), the acoustic modes satisfying this condition have very small axial wavenumbers $q_z$, resulting in acoustic frequencies close to their cutoff values. As shown in the Supplementary Material, the Brillouin shift can be calculated from the acoustic velocity, the cladding radius $R$, and the acoustic mode indices $l$ and $m$.

Multiple acoustic modes may satisfy this condition, leading to a series of resonances in intra-modal forward scattering. Notably, the same acoustic mode mediates both Stokes and anti-Stokes processes, such that the occurrence of one process necessarily implies the other. This feature leads to fundamentally different gain characteristics compared to backward Brillouin scattering \cite{kharel2016}.

\subsection{Optical and Acoustic Modes in a Fiber}
\begin{figure}[htbp]
  \centering
  \includegraphics[width=0.57\linewidth]{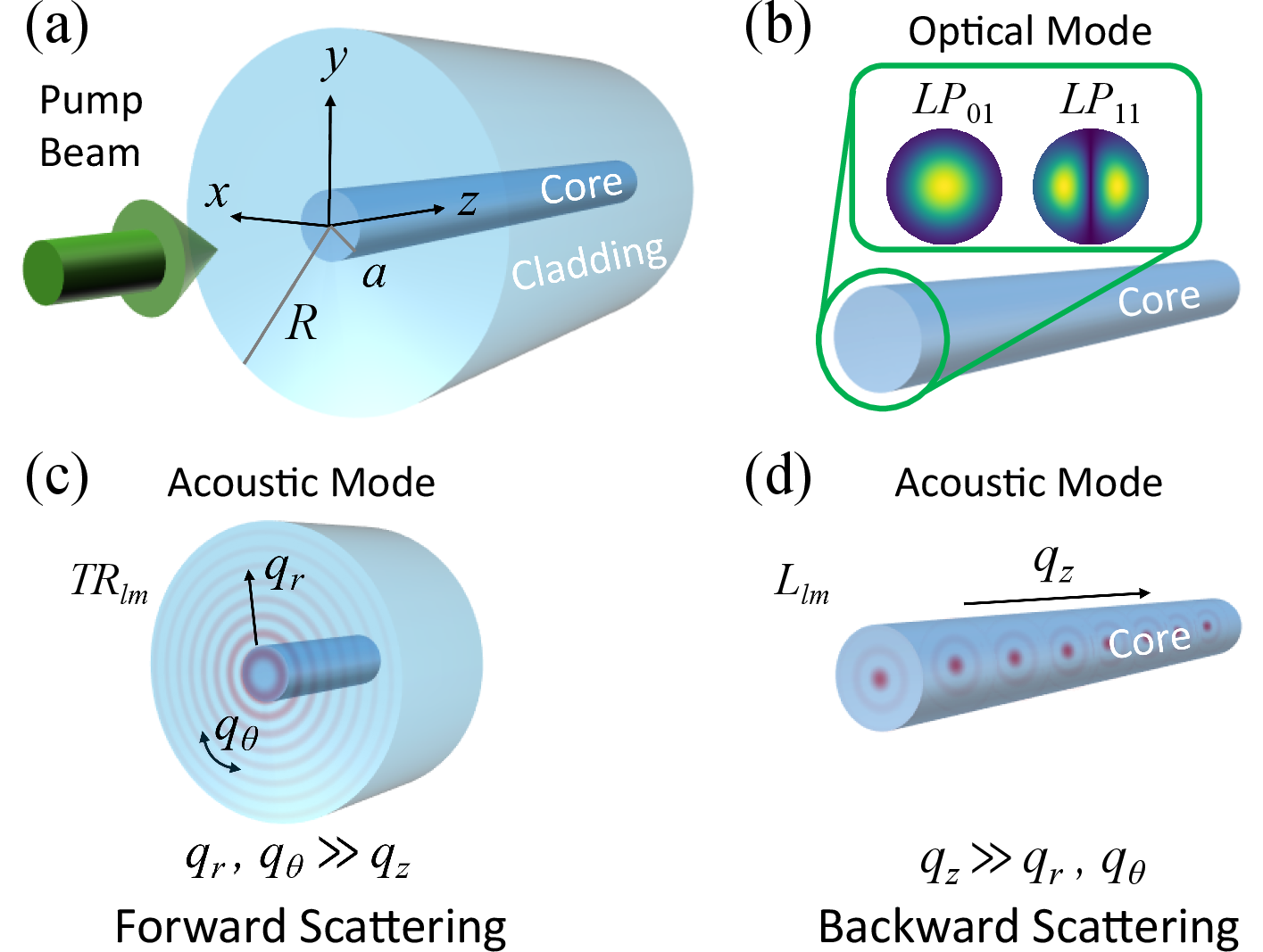} 
  \caption{(a) Schematic illustration of the optical fiber used in the experiment. A pump beam is launched into a few-mode fiber. (b) Optical modes supported in the fiber core, namely the $LP_{01}$ and $LP_{11}$ modes. (c) Dominant acoustic mode in forward Brillouin scattering, corresponding to a torsional–radial mode $TR_{lm}$ with a large transversal wavevector $q_r, q_\theta \gg q_z$. (d) Dominant acoustic mode in backward Brillouin scattering, corresponding to a longitudinal mode $L_{lm}$ with a large axial wavevector $q_z \gg q_r, q_\theta$.}
  \label{fig:optacomode}
\end{figure}

Figure~\ref{fig:optacomode}(a) shows the cartoon of the optical fiber system used in the experiment. A step-index optical fiber consists of a core with uniform refractive index and radius $a$, surrounded by a cladding with a lower refractive index and radius $R$ \cite{yariv_yeh_2007,okamoto2021fundamentals}. Optical modes in the fiber are conveniently described using linearly polarized (LP) modes, denoted $LP_{lm}$, where $l$ and $m$ represent the azimuthal and radial mode orders, respectively.

Unlike optical modes, which are primarily confined to the core, acoustic modes extend into both the core and the cladding regions \cite{wolff2021brillouin}. Acoustic waves may propagate along the fiber axis and in the transverse direction, and can be either longitudinal or shear in nature. The acoustic modes participating in different Brillouin scattering processes are summarized in Table~\ref{tab:2}.  
\begin{table}[h]
    \centering
    \caption{Characteristics of the acoustic modes for different Brillouin scattering}
    \label{tab:2}
        \begin{tabular}{cccccc}
            \hline
            \makecell{Brillouin \\Scattering \\ Direction} &\makecell{Acoustic\\wavenumber} & \makecell{Main\\Propagation\\Direction}&\makecell{Propagation \\Region} &\makecell{Propagation\\ Mode}&\makecell{Spatial\\Mode}\\
            \hline
            Backward&$q_z\gg q_r,q_\theta$ &$z$-direction& Core&Longitudinal& $L_{lm}$\\
            \hline
            Forward&$q_z\ll q_r,q_\theta$ &$xy$-direction& \makecell{Core and \\ Cladding}&\makecell{Shear and\\Longitudinal}& $TR_{lm}$ \\
            \hline
        \end{tabular}
\end{table}

Backward Brillouin scattering requires a large axial acoustic wavenumber $q_z$. When $q_z \gg q_x, q_y$, the acoustic wave is weakly guided along the fiber axis and confined within the core, forming discrete longitudinal guided modes $L_{lm}$ analogous to optical fiber modes \cite{dong2010formulation}. Owing to symmetry, modes $L_{0m}$ couple to intra-modal scattering with $LP_{01}$, whereas modes $L_{1m}$ participate in inter-modal scattering between $LP_{01}$ and $LP_{11}$.

In forward Brillouin scattering, the required axial wavenumber is small, $q_z \ll q_x, q_y$, and the acoustic wave propagates predominantly in the transverse direction. Both longitudinal and shear acoustic waves with velocities $v_{\mathrm{L}}$ and $v_{\mathrm{S}}$ can contribute. The corresponding radial wavenumbers are approximately $q_{r,\mathrm{L}} \simeq \Omega/v_{\mathrm{L}}$ and $q_{r,\mathrm{S}} \simeq \Omega/v_{\mathrm{S}}$. These transverse acoustic modes form torsional–radial modes, denoted $TR_{lm}$. Intra-modal forward scattering is dominated by $TR_{2m}$ modes, whereas inter-modal forward scattering primarily involves $TR_{1m}$ and $TR_{3m}$ modes.

\subsection{Brillouin Gain}
Electrostriction serves as the primary driving mechanism for generating acoustic waves in dielectric media \cite{PhysRevB.86.224304}. The same mechanism is responsible for the Brillouin scattering, where the input photon is scattered by an acoustic mode. The coupling strength between the optical and acoustical modes allows the qualitative determination of the gain in Brillouin scattering. 

For the optical fields $\vec{E}_{1}$ and $\vec{E}_{2}$, expressed in cylindrical coordinates as,

\begin{align}
    \vec{E}_{1}(r,\theta,z,t)&=H_1\phi_{1}(r,\theta) e^{i(k_{1z} z-\omega_1 t)}\hat{e}_1+\text{c.c.}\\
    \vec{E}_{2}(r,\theta,z,t)&=H_2\phi_{2}(r,\theta)e^{i(k_{2z} z-\omega_2 t)}\hat{e}_2+\text{c.c.},
\end{align}
where $H_i$ is the electric field amplitude and $\phi_i(r,\theta)$ is the spatial distribution of the field, normalized as $\iint |\phi_i(r,\theta)|^2 rdrd\theta=1$. $\hat{e}_i$ is the polarization vector.  The electrostriction force $\vec{F}$ is written as~\cite{1130012844551500709}.
\begin{equation}
\vec{F}=\frac{1}{4nc}\vec{f}(r,\theta)P\:\exp{i(q_zz-\Omega t)},
\end{equation}
where $P=2n\epsilon_0cA_1A_2^*$ is the optical beat power and $\vec{f}$ is the generalized force density, whose component is given by,
\begin{align}
f_i
&= n^4 \sum_{k,l}\left[\frac{\partial}{\partial x}\Bigl(p_{ixkl}\phi_1\phi_2^*(\hat{e}_1\!\cdot\!\hat{e}_k)(\hat{e}_2\!\cdot\!\hat{e}_l)\Bigr)+ \frac{\partial}{\partial y}\Bigl(p_{iykl}\phi_1\phi_2^*
(\hat{e}_1\!\cdot\!\hat{e}_k)(\hat{e}_2\!\cdot\!\hat{e}_l)\Bigr)\right.\notag\\
&\quad\left.+ i q_z\Bigl(p_{izkl}\phi_1\phi_2^*(\hat{e}_1\!\cdot\!\hat{e}_k)(\hat{e}_2\!\cdot\!\hat{e}_l)\Bigr)\right]
\end{align}
where $i,k,l\in \{x,y,z\}$ and $p_{ijkl}$ is the photoelastic tensor.

For the backward scattering, $q_z$ is large, and $f_z$ is the primary force component.  While for the forward scattering, the transverse components $f_x$ and $f_y$ (equivalently, $f_r$ and $f_\theta$) become dominant. The overlap between the acoustic mode and the driving electrostriction force is quantified by the coupling integral:
\begin{equation}
Q=\int \vec{u}\cdot \vec{f}dA,
\end{equation}
where $\vec{u}(r,\theta)$ is the spatial distribution of the acoustic displacement vector, also normalized as $\iint\vec{u}\cdot\vec{u}^* rdrd\theta=1$. 
The Brillouin gain $G(\Omega)$ and Brillouin gain coefficient $G_\text{B}$ can then be expressed in terms of the coupling integral $Q$ as \cite{layosh2025forward}
\begin{equation}
G(\Omega)=\frac{kQ^2}{8n^2c\rho\Gamma\Omega}\frac{i\Gamma/2}{\Omega_{lm}-\Omega+i\Gamma/2} ,\quad G_\text{B}=\frac{kQ^2}{8n^2c\rho\Gamma\Omega_{lm}},
\label{eq:TheoryGain}
\end{equation}
where $k = 2\pi/\uplambda$ is the optical wavenumber in vacuum, $\rho$ is the material density, $\Gamma$ is the acoustic damping rate, and $\Omega_{lm}$ is the phase-matched acoustic frequency.

\subsection{Spontaneous Brillouin scattering}

Spontaneous Brillouin scattering occurs when an incident pump beam interacts with thermally excited acoustic phonons in a medium, resulting in the generation of frequency-shifted scattered light. Unlike stimulated Brillouin scattering, which requires a sufficiently strong pump to initiate and amplify the scattering process, spontaneous Brillouin scattering arises solely from thermal fluctuations and is present even in the absence of any externally seeded Stokes wave.

In the case of backward Brillouin scattering, the ratio between the incident pump power $P_\text{P}$ and the spontaneously scattered Stokes power $P_\text{S}$ is given by \cite{PhysRevA.42.5514}

\begin{equation}
\frac{P_\text{S}}{P_\text{P}}=\frac{\omega_\text{P}G_\text{B}k_\text{B}   T L\Gamma }{4\Omega_{lm}}e^{(G_\text{B}P_\text{P}L/2)}\left[I_0(G_\text{B}P_\text{P}L/2)-I_1(G_\text{B}P_\text{P}L/2)\right]
\label{eq:Spontaneous BS}
\end{equation}
where $\omega_\text{P}$ is the pump frequency, $k_\text{B}$ is the Boltzmann constant, $T$ is the temperature, $L$ is the interaction length. $I_j$ denotes the modified Bessel function of the first kind of order $j$.

In intra-modal forward Brillouin scattering, the same acoustic mode mediates both the Stokes and anti-Stokes processes. Consequently, the powers of the Stokes and anti-Stokes waves become nearly equal, and the net amplification exhibits a linear dependence on the pump power. In this case, the ratio of the spontaneously scattered power $P_\text{S}$ to the incident pump power $P_\text{P}$ is given by \cite{kharel2016}
\begin{equation}
\frac{P_\text{S}}{P_\text{P}}=\frac{\omega_\text{P} G_\text{B} k_\text{B} T L \Gamma}{4 \Omega_{lm}}.
\label{eq:FWsp}
\end{equation}

\section{Experiment}
The Brillouin gain medium used in this experiment is a 1-km-long single-mode fiber (SMF-28 Ultra), originally designed for operation in the optical communication band ($\uplambda = 1260-1625$~nm). The specifications of this step-index optical fiber are listed in Table~\ref{tab:3}. At the laser wavelength used in this experiment, $\uplambda = 1064$~nm, the V-number of the fiber is estimated to be $V = \frac{2\pi a}{\uplambda}\sqrt{n_1^2 - n_2^2} = 3.097$, indicating that the fiber supports two guided modes at this wavelength. Specifically, the supported optical modes are the fundamental mode $LP_{01}$ and the first higher-order mode $LP_{11}$.

\begin{table}[h!]
    \centering
    \caption{Specifications of the SMF-28 Ultra fiber. The photoelastic constants are taken from D. Donadio \emph{et~al.}~\cite{PhysRevB.70.214205}.}
    \label{tab:3}
        \begin{tabular}{|l|c|}
            \hline
            \textbf{Parameter} & \textbf{Value} \\
            \hline
            Fiber Type & SMF-28 Ultra \\
            \hline
            Radius (Core, Cladding)& $a=4.2$~$\upmu$m, $R=62.5$~$\upmu$m\\
            \hline
            Refractive Index & $n_\text{core}=1.4730$, $n_\text{clad}=1.4677$ \\
            \hline
            Effective Refractive Index ($LP_{01}$, $LP_{11}$) & $n_\mathrm{01}=1.4712$, $n_\mathrm{11}=1.4687$\\
            \hline
            Density  & \(\rho=2203~\text{kg}/\text{m}^3\) \\
            \hline
            Longitudinal Acoustic Velocity & $v_\mathrm{core,L}=5720$~m/s, $v_\mathrm{clad,L}=5964$~m/s\\
            \hline
            Shear Acoustic Velocity & $v_\mathrm{clad,S}=3739$~m/s \\
            \hline
            Photoelastic Constant (in Voigt notation) & $p_{11}=0.125$, $p_{12}=0.27$, $p_{44}=\frac{p_{11}-p_{12}}{2}$  \\
            \hline
        \end{tabular}
\end{table}

\begin{figure}[htbp]
  \centering
  \includegraphics[width=0.76\linewidth]{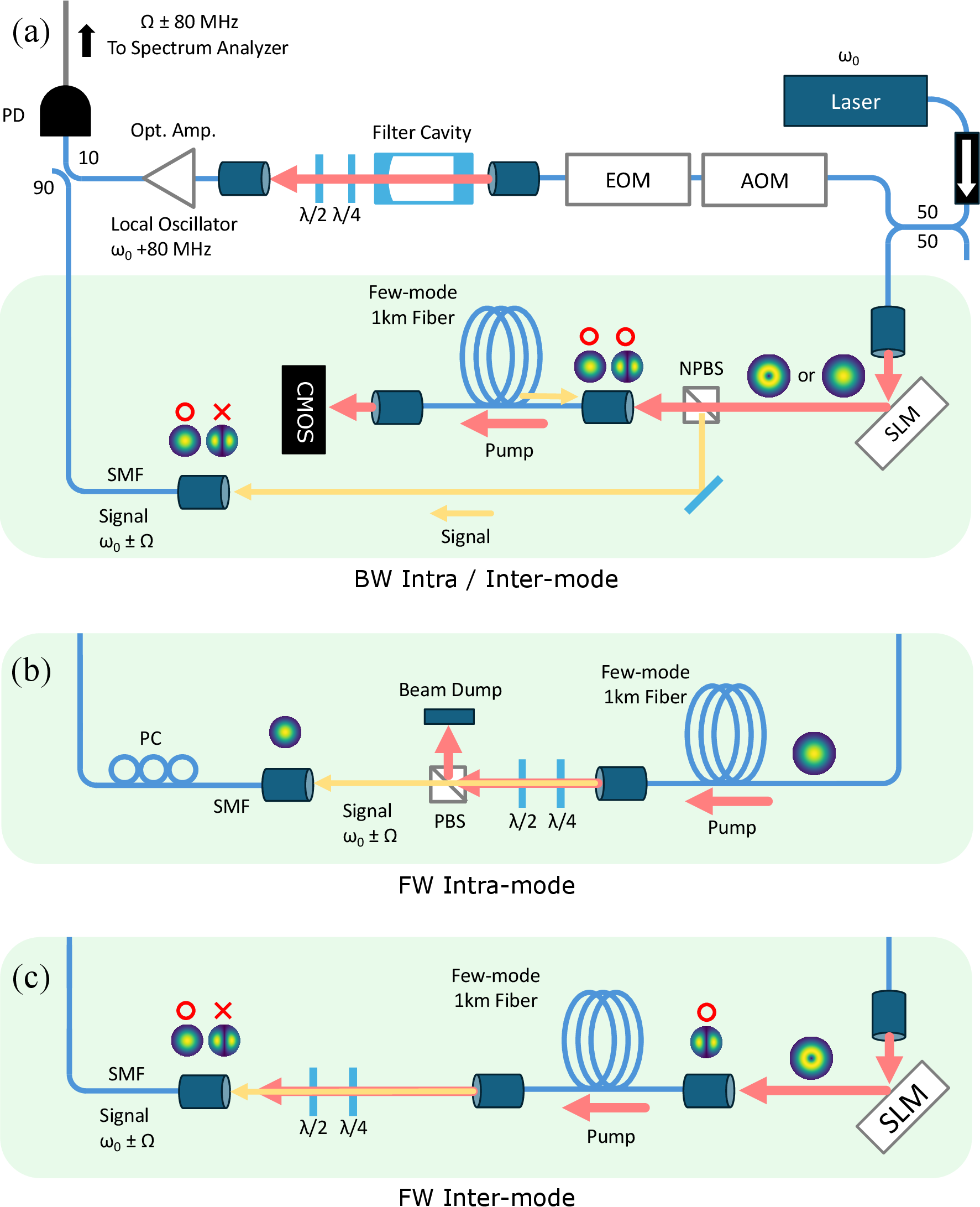}
  \caption{
(a) Experimental setup for measuring backward Brillouin scattering. The laser beam is split into two paths. (Top) One path is used to generate the local oscillator (LO) with a frequency shift introduced by an acousto-optic modulator (AOM) and an electro-optic modulator (EOM). A filter cavity removes unwanted sideband signals, followed by polarization control and optical amplification. (Bottom) The other path is shaped by a spatial light modulator (SLM) to generate the desired spatial profile of the pump beam. The backward-scattered light is separated using a non-polarizing beam splitter (NPBS) and coupled into a single-mode fiber to suppress non-$LP_{01}$ components. The spatially filtered signal is then mixed with the LO beam. The resulting heterodyne signal is detected by a photodetector and analyzed using a spectrum analyzer.
(b) Experimental setup for detecting intra-modal forward Brillouin scattering. Polarization-controlling waveplates and a polarizing beam splitter (PBS) are used to suppress the pump beam and extract only the orthogonally polarized component of the scattered light. This signal is subsequently coupled into a single-mode fiber (SMF) for heterodyne detection.
(c) Experimental setup for detecting inter-modal forward Brillouin scattering. A Laguerre--Gaussian beam is used to excite the $LP_{11}$ mode in an SMF-28 fiber. The scattered signal is collected via free-space coupling and filtered by a single-mode fiber to isolate the $LP_{01}$ component.
}
  \label{fig:setup}
\end{figure}

The optical setup used for the backward scattering measurement is shown in Fig.~\ref{fig:setup}(a). The output of a single-mode narrowband laser (Coherent Mephisto S) at $\uplambda=1064$~nm is split into two paths to generate the pump and local oscillator (LO) beams. The pump beam excites the few-mode fiber, and the backward Brillouin signal is subsequently mixed with the LO beam for heterodyne detection.  

The maximum pump beam power available is approximately 200~mW. The spatial mode of the pump beam is controlled using a spatial light modulator (SLM: Santec SLM-200). To excite the $LP_{01}$ mode, the output of a single-mode fiber is directly coupled into the few-mode fiber. For excitation of the $LP_{11}$ mode~\cite{Bruning:15}, a Laguerre-Gaussian (LG) beam is generated using the SLM, with a spatial-mode conversion efficiency of approximately 40\%.  The alignment of the pump beam into the 1-km-long few-mode fiber is optimized by monitoring the output beam profile with a CMOS camera and/or by maximizing the transmitted optical power.  

The frequency of the LO beam is shifted by an acousto-optic modulator (AOM) and an electro-optic modulator (EOM). To suppress unwanted frequency components generated by the EOM, the beam is passed through a filter cavity with a linewidth of 10~MHz.  The transmission of the filter cavity is monitored using a photodiode, and a PID feedback loop is employed to lock the cavity to one of the modulated sidebands, thereby generating a single-frequency LO beam for heterodyne detection.

For backward scattering measurements, the signal light is extracted using a non-polarizing beam splitter placed at the input of the few-mode fiber and subsequently coupled into a single-mode detection fiber.  This detection fiber spatially filters the signal, eliminating residual non-$LP_{01}$ optical modes. The filtered signal is then combined with the LO beam and detected using a fast photodetector (New Focus 1554-B). The resulting heterodyne signal is amplified and recorded with a spectrum analyzer (Agilent E4405B).  The absolute signal power is determined using a calibrated EOM by comparing the measured signal amplitude with that of the calibrated EOM sidebands.  

The experimental setups for intra-modal and inter-modal forward scattering measurements are shown in Fig.~\ref{fig:setup}(b) and Fig.~\ref{fig:setup}(c), respectively. The generation of the LO beam is identical to that used in the backward scattering measurement. For intra-modal forward scattering, the scattered light component with polarization orthogonal to that of the pump beam is selectively detected. The output of the few-mode fiber is again coupled into a single-mode detection fiber for heterodyne detection. 

For inter-modal forward scattering, in which an LG beam is used to excite the $LP_{11}$ mode, residual $LP_{11}$ components in the scattered signal are suppressed by the single-mode detection fiber to achieve a high signal-to-noise ratio.    

\section{Results}
\subsection{Backward Scattering}  

\begin{figure}[htbp]
  \centering
  \includegraphics[width=0.65\linewidth]{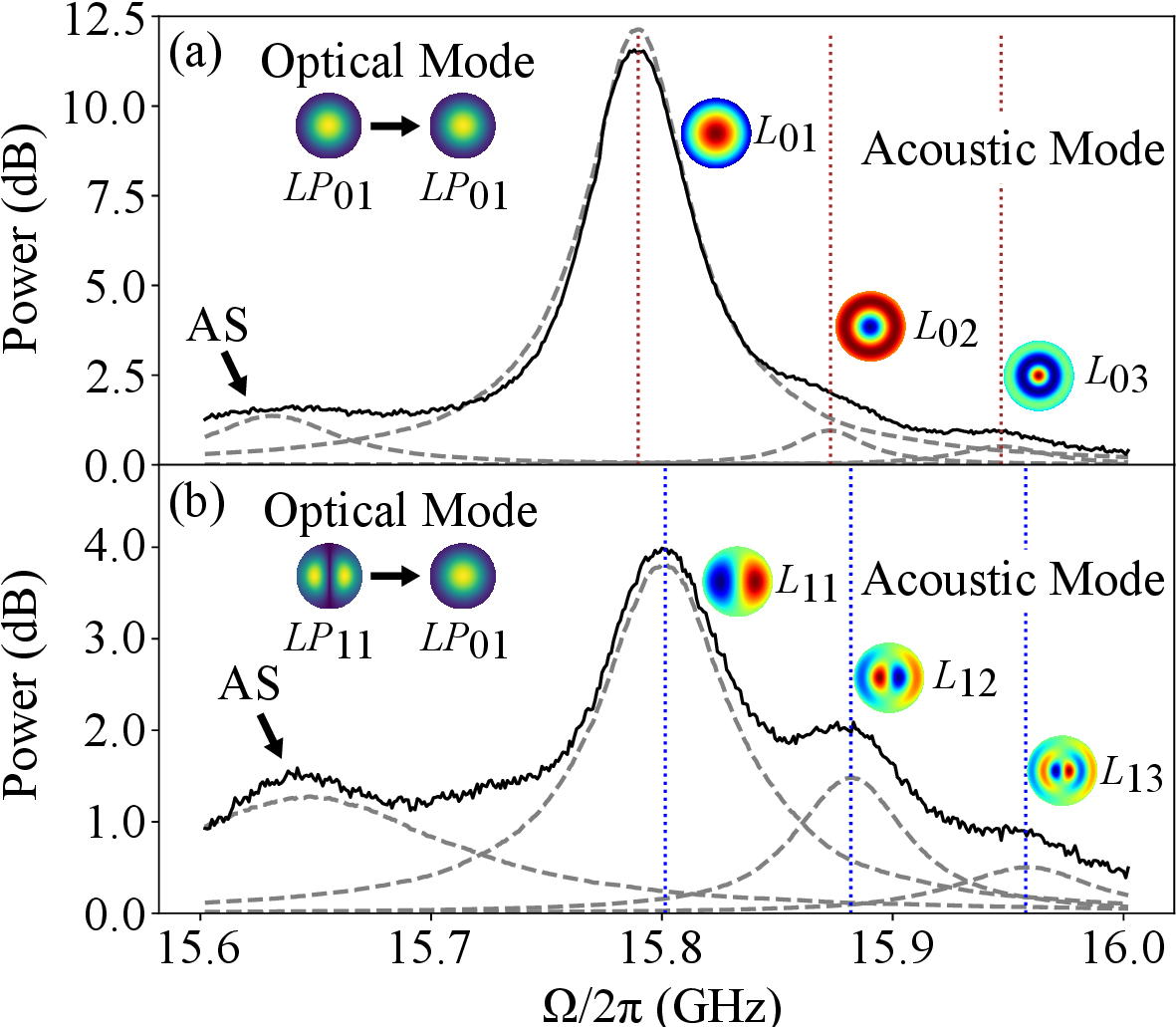} 
  \caption{Backward scattering spectra obtained from heterodyne measurements. (a) Intra-modal scattering spectrum ($LP_{01}\rightarrow LP_{01}$). Three Stokes resonances corresponding to distinct acoustic modes, $L_{01}$, $L_{02}$, and $L_{03}$, are observed around $\Omega/2\pi \approx 16$~GHz. The reference level of the vertical axis (0.0~dB) is set to the electrical noise floor. (b) Inter-modal scattering spectrum ($LP_{11}\rightarrow LP_{01}$). Three resonances associated with acoustic modes of azimuthal order $l=1$, namely $L_{11}$, $L_{12}$, and $L_{13}$, are observed. Slight shifts in the resonance frequencies originate from differences in the acoustic phase velocities. In both spectra, weaker anti-Stokes (AS) signals are also observed, as indicated in the figure.}
  \label{fig:BBSmodes}
\end{figure}

Figure~\ref{fig:BBSmodes}(a) shows the Stokes signal obtained from intra-modal backwards Brillouin scattering, in which both the input and output optical modes are $LP_{01}$.  The measurement was performed using an input pump power of $P=16.0$~mW. Because a large acoustic wavenumber is required for backward scattering, the corresponding Brillouin shift appears around $\Omega/2\pi\approx 16$~GHz.The spectrum shown by the solid line contains the scattering contributions from three distinct acoustic modes. 

The measured spectrum is fitted using the noise-initiated Brillouin scattering model described in Eq.~\ref{eq:Spontaneous BS}. The individual spectral components corresponding to each acoustic modes, as extracted from the fitting procedure, are shown as dashed lines. 
It should be noted that the linewidths obtained directly from the gain-spectrum fitting tend to be larger than the intrinsic spectral linewidths, particularly in the high-gain regime. This discrepancy arises from the exponential terms included in the fitting model of Eq.~\ref{eq:Spontaneous BS}, which broaden the apparent linewidth in the fitted spectra. To accurately determine the intrinsic linewidth, the measured spectra are therefore independently fitted using a Lorentzian function. The linewidths obtained from the gain-spectrum fitting are typically larger by a factor of approximately 1.5 compared with those extracted from the Lorentzian fits.

From the fitting analysis, the Brillouin gain coefficient $G_\text{B}$, Brillouin shift frequency $\Omega$, and Brillouin linewidth $\Gamma$ are extracted. The acoustic phase velocity $v_\text{p}$  of each mode is subsequently determined from the corresponding Brillouin shift frequency $\Omega$. A summary of the parameters obtained from the fitting and subsequent calculations is provided in Table~\ref{tab:4}.

\begin{table}[h]
\centering
\caption{Calculated parameters obtained from the data for backward scattering. }
\begin{tabular}{c|cccccc}
\hline
Optical Mode & \makecell{Acoustic \\Mode \\$L_{lm}$}& \makecell{$G_\mathrm{B}$ \\ (W$^{-1}$km$^{-1}$)\\(Theory)}&\makecell{$G_\mathrm{B}$\\ (W$^{-1}$km$^{-1}$)\\ \:}&\makecell{$\Omega/2\pi$\\(GHz)}& \makecell{$\Gamma/2\pi$\\(MHz)}& \makecell{$v_\mathrm{p}$\\(m/s)} \\
\hline
Intra-modal&$L_{01}$& 258&160 & 15.790 & 30 & 5,710 \\
$LP_{01}\rightarrow LP_{01}$&$L_{02}$& 0.07&8.6 &15.873 & 36 & 5,740 \\
&$L_{03}$ & 0.12&6.9 & 15.947 & 45 & 5,767 \\
\hline
Inter-modal&$L_{11}$ & 54& 27&15.801 & 56 & 5,719 \\
$LP_{11}\rightarrow LP_{01}$&$L_{12}$ & 2.7&9.3& 15.882 & 52 & 5,749 \\
&$L_{13}$& 0.84&2.5& 15.958 & 72 & 5,775 \\
\hline
\end{tabular}
\label{tab:4}
\end{table}

Due to azimuthal symmetry, the fundamental optical mode $LP_{01}$ can couple only to longitudinal acoustic modes of the form $L_{0l}$. The three modes observed in the spectrum are $L_{01}$, $L_{02}$, and $L_{03}$ and their mode profiles are shown in the inset of Fig.~\ref{fig:BBSmodes}(a).  The fundamental longitudinal acoustic mode $L_{01}$ couples efficiently to the $LP_{01}$ optical mode owing to their similar spatial mode profiles and the resulting large mode overlap factor $Q$. Additionally, for backward scattering, the strong confinement of the acoustic mode within the fiber core leads to a small effective mode volume, further enhancing the opto-acoustic coupling. As a result, a large Brillouin gain coefficient $G_\text{B}=160$~W$^{-1}$km$^{-1}$ is obtained for the fundamental acoustic mode. In contrast, the coupling strength decreases rapidly for higher-order acoustic modes $L_{02}$ and $L_{03}$, which are nearly orthogonal to the $LP_{01}$ optical mode.  A small hump observed on the lower-frequency side of the spectrum originates from the anti-Stokes process and is strongly suppressed relative to the Stokes signal. 

In backward scattering, the inter-polarization scattering is typically forbidden in isotropic media due to symmetry considerations. Consequently, we focus on the spatial inter-modal scattering.  Inter-modal backward Brillouin scattering measurements are performed by exciting the fiber with a Laguerre–Gaussian (LG) pump beam, which can be regarded as a superposition of $LP_{11}$ modes oriented along the $x$ and $y$ directions, while monitoring the scattered signal in the $LP_{01}$ mode. 

The obtained spectrum is shown in Fig.~\ref{fig:BBSmodes}(b), obtained with an input pump power of $P=14.5$~mW. As in the intra-modal case, resonances associated with three acoustic modes are observed.  For this optical configuration, the acoustic modes must exhibit odd azimuthal symmetry in order to yield a nonzero mode overlap $Q$.  Accordingly, the resonances are identified as the acoustic modes $L_{11}$, $L_{12}$, and $L_{13}$. The corresponding acoustic mode profiles are shown in the inset of Fig.~\ref{fig:BBSmodes}(b). The slight variation in the Brillouin shift $\Omega$ for $L_{1m}$, with respect to $L_{0m}$ observed in the intra-modal scattering, reflects differences in the corresponding acoustic phase velocities. A summary of the parameters obtained from the fitting analysis and subsequent calculations is provided in Table~\ref{tab:4}.

\subsection{Forward Scattering}
\begin{figure}[!h]
\centering
\includegraphics[width=1\linewidth]{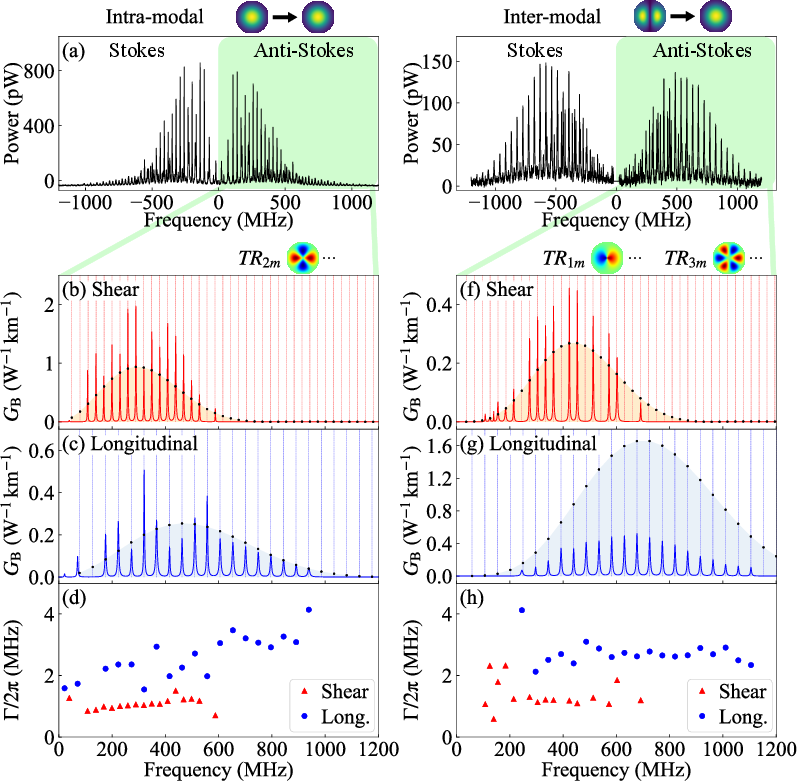}
 \caption{Forward scattering spectra for (a-d) intra-modal and (e-h) inter-modal scattering. For both scattering processes [see (a) and (e)],  nearly identical series of resonances for Stokes and anti-Stokes processes are observed, ranging from a few MHz to a GHz. (b) and (f) show a comparison of the Brillouin gain calculated from data (solid line) and from calculation (dots) for shear acoustic waves. The orange shade shows the variation of the calculated gain. (c) and (g) show the gain comparison described above for the longitudinal acoustic waves.  The blue shaded area shows the calculated gain.  (d) and (h) show the variation of the Brillouin linewidth $\Gamma$ for shear (red triangle) and longitudinal (blue circle) acoustic modes observed. }
 \label{fig:FW}
\end{figure}

Figure~\ref{fig:FW} presents the forward Brillouin scattering spectra for (a--d) intra-modal and (e--h) inter-modal scattering configurations. The horizontal axis corresponds to the frequency shift of the scattered light: negative frequencies indicate Stokes components, while positive frequencies correspond to anti-Stokes components.  In the weak-gain regime, as in the present measurements, the Stokes and anti-Stokes components exhibit nearly identical gain, resulting in spectra that are symmetric about zero frequency.

In forward Brillouin scattering, the dominant acoustic modes participating in the interaction are cladding modes, for which the boundary conditions are defined by a large diameter of the fiber cladding. As a result, the corresponding Brillouin frequencies are significantly lower than those observed in backward scattering. Additionally, numerous resonances are found in the spectra.  This is owing to the existence of a series of ``optical-like'' acoustic modes, which can satisfy the phase matching condition, $v=c/n$. 

In the intra-spatial-modal scattering configuration, acoustic modes belonging to the $TR_{2m}$ family are observed. In this case, scattering occurs between orthogonally polarized $LP_{01}$ optical modes, as illustrated in the experimental setup shown in Fig.~\ref{fig:setup}(b). In contrast, inter-modal forward Brillouin scattering probes acoustic modes from the $TR_{1m}$ and $TR_{3m}$ families, with the optical pump and probe propagating in different spatial modes ($LP_{01}$ and $LP_{11}$), as shown in Fig.~\ref{fig:setup}(c).

For both intra- and inter-modal scattering, the measured spectra are fitted with Lorentzian functions, and the corresponding Brillouin gain spectra are calculated using Eq.~\ref{eq:FWsp}. Each resonance peak is classified as either a longitudinal or a shear acoustic mode based on its center frequency. The two mode families are plotted separately: shear modes in Fig.~\ref{fig:FW}(b,f) and longitudinal modes in Fig.~\ref{fig:FW}(c,g). Theoretical Brillouin gain spectra, calculated using the $TR_{2m}$ acoustic modes for intra-modal scattering and the $TR_{1m}$ modes for inter-modal scattering, are overlaid on the experimental data as black dots and shaded regions.

Both experimental measurements and theoretical calculations demonstrate that the spectral envelope of the inter-modal scattering is higher than that of the intra-modal scattering, while maintaining good agreement in spectral shape. As discussed above, the acoustic frequencies of the $TR_{1m}$ and $TR_{3m}$ modes overlap in forward scattering. Consequently, the measured inter-modal spectrum represents the combined contribution of these two acoustic mode families. An example of two modes with closely spaced frequencies ($\sim 423~\mathrm{MHz}$) is listed in Table~\ref{tab:5}. While the shear-mode gains of the $TR_{1m}$ and $TR_{3m}$ modes are nearly identical, the longitudinal-mode gain of the $TR_{3m}$ family is significantly smaller. The overlap of the shear-mode responses from the $TR_{1m}$ and $TR_{3m}$ modes therefore leads to an enhanced effective gain in the inter-modal scattering spectra, as observed in Fig.~\ref{fig:FW}.

The linewidths $\Gamma$ extracted from each resonance are plotted in Fig.~\ref{fig:FW}(d, h), with red triangles and blue circles denoting shear and longitudinal acoustic modes, respectively. The measured spectra reveal that resonances associated with longitudinal acoustic modes consistently exhibit broader linewidths than those associated with shear modes. A similar behavior has been reported in forward Brillouin scattering studies for sensing applications~\cite{bernstein2023tensor}, where the effect is attributed to interactions with materials outside the fiber cladding. In the present experiment, the polymer coating surrounding the fiber modifies the boundary conditions for longitudinal acoustic waves, leading to increased acoustic damping, while exerting only a minor influence on shear acoustic modes.

\begin{table}[h]
\centering
\caption{Calculated parameters obtained from the data for forward scattering. The symbols S and L in the acoustic mode denote the shear and longitudinal modes, respectively.}
\begin{tabular}{c|cc@{}c@{}cc@{}c@{}cc}
\hline
Optical Mode & \makecell{Acoustic \\Mode\\$TR_{lm}$}& \makecell{$G_\mathrm{B}$ \\ (W$^{-1}$km$^{-1}$)\\(Theory)}&&\makecell{$G_\mathrm{B}$\\ (W$^{-1}$km$^{-1}$)\\\:}&\makecell{$\Omega/2\pi$\\(MHz)\\(Theory)}&& \makecell{$\Omega/2\pi$\\(MHz)\\\:}& \makecell{$\Gamma/2\pi$\\(MHz)} \\
\hline
Intra-modal&$TR_{2,9}$ (S)& 0.94 && 2.0 $\pm$ 0.5  & 289.0 && 289 &1.1\\
$LP_{01}\rightarrow LP_{01}$&$TR_{2,10}$ (L)& 0.25&& 0.28 $\pm$ 0.08 &510.1 &&511  & 2.7 \\
\hline
\multirow{4}{*}{\makecell{Inter-modal \\$LP_{11}\rightarrow LP_{01}$}}& $TR_{1,14}$ (S) & 0.27&\multirow{2}{*}{$\left.\rule{0pt}{1.8em}\right\}$} 
& \multirow{2}{*}{0.46 $\pm$ 0.14} & 423.1 &\multirow{2}{*}{$\left.\rule{0pt}{1.8em}\right\}$} & \multirow{2}{*}{423} & \multirow{2}{*}{1.2} \\
& $TR_{3,13}$ (S) & 0.27&&  & 422.3 &&  &  \\
& $TR_{1,14}$ (L) & 1.66 &\multirow{2}{*}{$\left.\rule{0pt}{1.8em}\right\}$}& \multirow{2}{*}{0.52 $\pm$ 0.13} & 677.2 &\multirow{2}{*}{$\left.\rule{0pt}{1.8em}\right\}$} & \multirow{2}{*}{677}& \multirow{2}{*}{2.6} \\
& $TR_{3,13}$ (L) & 0.07&&  & 675.8 &&  &  \\
\hline
\end{tabular}
\label{tab:5}
\end{table}

\section{Discussion}
In backward Brillouin scattering, regardless of whether the process is intra- or inter-modal, the fundamental acoustic mode exhibits a very strong coupling. In addition, scattering into a few higher-order acoustic modes with much weaker strengths is observed. This behavior originates from the fact that both the optical and acoustic waves form weakly guided modes confined within the fiber core, resulting in highly similar spatial mode profiles. While the fundamental acoustic mode has an almost complete spatial overlap with the optical mode, higher-order acoustic modes are essentially orthogonal to the optical mode, leading to a drastic reduction in the coupling strength. In the present measurements, a maximum Brillouin gain coefficient of $G_\mathrm{B} = 160~\mathrm{W^{-1}km^{-1}}$ is observed. 

From the measured Brillouin frequency shifts, the acoustic phase velocities can be calculated. As predicted by theory, the phase velocity increases for higher-order acoustic modes. Although the experimentally obtained Brillouin gains are slightly lower than the theoretical values, the overall agreement between experiment and theory is satisfactory.

In forward Brillouin scattering, in contrast to the backward case, the optical and acoustic modes are subject to different boundary conditions in the core and cladding, respectively, resulting in fundamentally different spatial mode profiles. Consequently, the coupling is generally weaker; however, no mode is completely canceled, and a large number of weak couplings appear as multiple resonances over a broad frequency range. 

Unlike backward scattering, both shear and longitudinal acoustic modes are clearly resolved in the forward-scattering spectra. The observed resonance frequencies are in excellent agreement with theoretical predictions, and Brillouin gains are also highly reproducible and consistent with theory. Although the overall gain amplitude differs by approximately a factor of two in some cases, the spectral envelopes show very good agreement. A maximum gain coefficient of $G_\mathrm{B} = 2.0~\mathrm{W^{-1}km^{-1}}$ is observed, which is in close agreement with previously reported experimental value $G_\text{B}=1.2$~W$^{-1}$km$^{-1}$ in a polarization-maintaining fiber for the $TR_{2m}$ mode at 169~MHz~\cite{bashan2021forward}.  For comparison, when forward Brillouin scattering is exploited as a nonlinear-optical device and the spatial modes are matched, significantly larger gain coefficients have been reported, for example in photonic crystal structures ($G_\mathrm{B} \approx 1,500~\mathrm{W^{-1}km^{-1}}$) and in liquid-core fibers using carbon disulfide ($G_\mathrm{B} \approx 5,800~\mathrm{W^{-1}km^{-1}}$)~\cite{kang2009tightly, behunin2019spontaneous}. 

The quality factor of the acoustic modes in optical fibers is approximately $\Omega/\Gamma \sim 500$ for backward scattering, whereas it decreases to about $\Omega/\Gamma \sim 300$ in forward scattering. This reduction is attributed to the fact that, in forward scattering, the acoustic modes are cladding modes and are therefore subject to additional absorption losses associated with the fiber jacket, as reported in earlier studies~\cite{bernstein2023tensor}.

Regarding the technical aspects of the optical measurement setup, we demonstrate high--signal-to-noise-ratio spectroscopy by employing heterodyne detection in combination with spatial and polarization filtering. In addition, the spatially selective modification of the optical mode structure using a programmable spatial light modulator (SLM) serves as a versatile tool for intra-modal spectroscopy. A key advantage of this approach is its high degree of reconfigurability: the same optical setup can be readily adapted to fibers with different core diameters and numerical apertures, as well as to different spatial modes, simply by updating the SLM phase pattern and adjusting the focusing optics. In combination with single-mode-fiber-based fundamental-mode filtering, which enables extraction of the \(LP_{01}\) component from a multimode optical field, this scheme realizes spatial-domain mode demultiplexing without relying on dedicated mode-selective hardware such as photonic lanterns~\cite{leon2014mode,leon2010photonic} or integrated mode multiplexers~\cite{chen2014compact}.

In the present experiment, the acoustic modes are not externally driven; consequently, the measured spectra directly reflect the fundamental thermal-noise limit of the fiber. These characteristics provide quantitative metrics for evaluating optical fibers used in quantum memory and precision quantum measurement systems. Furthermore, beyond the specific acoustic modes observed here, the use of Laguerre--Gaussian beams carrying orbital angular momentum (OAM), together with the mode-selective detection technique, establishes the experimental groundwork required to manipulate and detect OAM states in both optical and acoustic modes. This foundation paves the way for fundamental studies, including those of chiral phonons, as well as for the realization of OAM-based acoustic quantum memories.

\section{Conclusion}
We have carried out a comprehensive measurement of spontaneous Brillouin scattering in a few-mode optical fiber, resolving forward and backward scattering in both intra- and inter-modal configurations. Heterodyne detection enabled quantitative determination of the Brillouin resonance frequencies, linewidths, and calibrated gain coefficients, in good agreement with theory. These results provide benchmark data for guided acoustic--optical interactions in few-mode fibers and support future mode-selective optomechanical and quantum applications.

\begin{backmatter}
\bmsection{Funding}
This work was supported by JST (Moonshot R\&D – Program) Grant Number (JPMJMS226C).
\bmsection{Disclosures}
The authors declare that there are no conflicts of interest related to this article.
\bmsection{Data Availability Statement}
Data underlying the results presented in this paper are not publicly available at this time but may be obtained from the authors upon reasonable request.
\bmsection{Supplemental document}
See Supplement 1 for supporting content.
\end{backmatter}

%%%%%%%%%%%%%%%%%%%%%%% References %%%%%%%%%%%%%%%%%%%%%%%%%
\bibliography{bibliography}

\end{document}